\documentclass[12pt]{article}
\usepackage{graphicx}
\usepackage{amsmath}
\usepackage{longtable}

\begin{document}

\makeatletter
\renewcommand*{\@cite}[2]{{#2}}
\renewcommand*{\@biblabel}[1]{#1.\hfill}
\makeatother

\title{Systematic Error of the Gaia DR1 TGAS Parallaxes from Data for the Red Giant Clump}
\author{George~A.~Gontcharov\thanks{E-mail: george.gontcharov@tdt.edu.vn}}

\maketitle

1. Department for Management of Science and Technology Development, Ton Duc Thang University, Ho Chi Minh City, Vietnam\\
2. Faculty of Applied Sciences, Ton Duc Thang University, Ho Chi Minh City, Vietnam

Key words: parallaxes, giant stars, Hertzsprung--Russell diagram, Galactic solar neighborhoods.

Received January 9, 2017

DOI: 10.1134/S1063773717060044

Abstract - Based on the Gaia DR1 TGAS parallaxes and photometry from the Tycho-2, Gaia, 2MASS, and WISE catalogues, we have produced a sample of about 100 000 clump red 
giants within about 800 pc of the Sun. The systematic variations of the mode of their absolute magnitude as a function of the distance, magnitude, and other parameters have been 
analyzed. We show that these variations reach 0.7 mag and cannot be explained by variations in the interstellar extinction or intrinsic properties of stars and by selection. The
only explanation seems to be a systematic error of the Gaia DR1 TGAS parallax dependent on the square of the observed distance in kpc: $0.18R^2$ mas. 
Allowance for this error reduces significantly the systematic dependences of the absolute magnitude mode on all parameters. This error reaches 0.1 mas within 800 pc
of the Sun and allows an upper limit for the accuracy of the TGAS parallaxes to be estimated as 0.2 mas. A careful allowance for such errors is needed to use clump red giants as 
``standard candles''. This eliminates all discrepancies between the theoretical and empirical estimates of the characteristics of these stars and allows us to obtain the first 
estimates of the modes of their absolute magnitudes from the Gaia parallaxes:
$mode(M_H)=-1.49^m\pm0.04^m$, $mode(M_{K_s})=-1.63^m\pm0.03^m$, $mode(M_{W1})=-1.67^m\pm0.05^m$, $mode(M_{W2})=-1.67^m\pm0.05^m$, 
$mode(M_{W3})=-1.66^m\pm0.02^m$, $mode(M_{W4})=-1.73^m\pm0.03^m$, as well as the corresponding estimates of their de-reddened colors.

\newpage

\section*{INTRODUCTION}

The first results of the Gaia space project were presented in September 2016 (Gaia 2016). Among them there is the Gaia DR1 Tycho--Gaia astrometric solution (TGAS) catalogue 
(Michalik et al. 2015) with accurate parallaxes for more than two million stars from the Tycho-2 catalogue (H\o g et al. 2000).
These parallaxes were obtained by comparing the Gaia data with the coordinates of the same stars from the Hipparcos (van Leeuwen 2007) and Tycho-2 catalogues.
The coordinates, proper motions, phorometry in the broad visual $G$ band, and other data were also obtained for these stars.

The Gaia parallaxes are important for calibrating the characteristics of stars, their classification, and establishing the cosmic distance scale. Therefore, it
is important to estimate the possible parallax errors.

Lindegren et al. (2016) compared the TGAS and Hipparcos parallaxes for 86928 stars and found the median ``TGAS minus Hipparcos'' parallax difference
$\Delta\varpi$ (i.e., the difference of the parallax zero points) to be $-0.089\pm0.006$ milliarcseconds (mas). This quantity is different northward and southward of the
ecliptic: $-0.130\pm0.006$ and $-0.053\pm0.006$ mas, respectively; $\Delta\varpi$ increases and decreases with color index $(V-I)$ northward and southward of the ecliptic,
respectively. When analyzing the TGAS parallaxes for more than 90000 quasars, Lindegren et al. (2016) found the ``TGAS minus quasars'' difference $\Delta\varpi$ to be 
$-0.073\pm0.002$ and $+0.074\pm0.005$ mas northward and southward of the ecliptic, respectively, and the same trends as a function of the color as
those for Hipparcos stars. In addition, Lindegren et al. (2016) compared the TGAS parallaxes with the parallaxes of 141 Cepheids derived from the period--luminosity relation 
and found the median difference $\Delta\varpi=-0.016\pm0.023$ mas.

Jao et al. (2016) compared the TGAS parallaxes with the trigonometric parallaxes (those from Hipparcos and the ground-based ones) for 612 single stars within 25 pc of the Sun. 
They found $\overline{\Delta\varpi}=-0.24\pm0.02$ mas. However, because of their small number and low accuracy, the ground-based parallaxes barely affected this value. 
This difference has the same sign as the difference found by Lindegren et al. (2016). Jao et al. (2016) did not find any dependence of $\Delta\varpi$ on the magnitude, color, 
and proper motion. They found a difference $\Delta\varpi$ northward and southward of the ecliptic but with the sign opposite to the result of Lindegren et al. (2016):
$-0.17\pm0.03$ and $-0.32\pm0.04$ mas, respectively.

Stassun and Torres (2016) compared the TGAS parallaxes with the parallaxes of 111 eclipsing binary stars determined from their physical characteristics.
The accuracy of these parallaxes is, on average, $0.19$ mas. The authors found the ``TGAS minus eclipsing binary stars'' difference to be $\Delta\varpi=-0.25\pm0.05$ mas. 
They also found noticeable dependences of this difference on the stellar temperature, magnitude $V$, and the parallax itself strongly correlating with $V$. 
They pointed out that for distant stars, at a parallax less than 1 mas, $\Delta\varpi$ approaches zero.
In addition, they found a noticeable difference $\Delta\varpi$ northward and southward of the ecliptic: $-0.38\pm0.06$ and $-0.05\pm0.09$ mas, respectively. The authors
found the dependence $\Delta\varpi=-0.22-0.003\beta$ mas, where $\beta$ is the ecliptic latitude in degrees.

De Ridder et al. (2016) compared the TGAS parallaxes with the asteroseismic parallaxes derived from various data for the stars observed with the Kepler telescope. 
They found good agreement between the asteroseismic and TGAS parallaxes for 22 dwarfs and subgiants close to the Sun but a significant difference
for 938 red giants in the range of distances from 0.5 to 5 kpc. In the opinion of the authors, the asteroseismic parallaxes must be approximately an order
of magnitude more accurate than the TGAS parallaxes. Therefore, the authors deemed the difference found to be a manifestation of the TGAS parallax
errors, although they did not give their values. They pointed out that the parallax errors are more noticeable against the background of low values of the
parallaxes themselves, i.e., for such distant stars as red giants.

These studies show that the TGAS parallaxes can have systematic errors at a level of tenths of a milliarcsecond.

To reveal and analyze these errors, we can calculate the absolute magnitudes from the TGAS parallaxes for a sample of ``standard candles'', i.e., stars
with predictable and approximately identical luminosities, and then analyze the variations of the absolute magnitude typical for the sample as a function
of their distance and other parameters. To be more precise, for a sample of such stars we analyze the variations of the quantities in the equation 
$m=5-5\log(R)-M-A$, where $m$ is the apparent magnitude known from observations, $R$ is the distance from the TGAS parallax, $M$ is the absolute magnitude,
and $A$ is the interstellar extinction. Such variations can arise from (1) the natural variations of the sample composition, primarily from the age and metallicity
variations in the Galaxy -- they then affect $M$, (2) the sample incompleteness -- they then affect $M$, (3) the variations in extinction or uncertainty in the extinction
correction -- they then affect $A$, and (4) the systematic errors of the TGAS parallaxes -- they then affect $R$. In this study we estimate the influence of all
these factors and reach the conclusion about the value and character of the fourth of them, i.e., the possible systematic error of the TGAS parallaxes.

\section*{ORIGINAL DATA AND THE METHOD}

In this study we use fairly widespread high-luminosity stars, clump giants, as standard candles. These are stars after the passage of the giant branch
and the helium flash. They consist of an inert hydrogen envelope and a helium core, where the nuclear reactions of helium conversion into carbon
take place. The clump giants were described by Gontcharov (2008), who obtained a sample of 97348 such stars from the Tycho-2 catalogue, showed this
sample to be complete within several hundred pc of the Sun, and analyzed its distribution in space and kinematics.

Astraatmadja and Bailer-Jones (2016) obtained and justified the most probable distance estimates for TGAS stars, which are below designated
as $R_{mode}$. In particular, they took into account the Lutz-Kelker and Malmquist biases (Perryman 2009, pp. 208--212). $R_{mode}$ differ systematically from the
distances estimated from the simple formula $R_{par}=1/\varpi$, though insignificantly near the Sun. Bailer-Jones (2015) showed that although the $R_{mode}$ estimates
are closer to the true ones than $R_{par}$, they also have a large error at a low relative accuracy of the parallax $\sigma(\varpi)/\varpi$. A parallax with a large error
$\sigma(\varpi)/\varpi>0.35$ gives little information about the distance. Note that, according to the recommendations of the TGAS authors, the uncertainty of 0.3 mas that
describes the disregarded systematic parallax errors should be added to the formal parallax error specified by them. Then, the space $R<800$ pc, in which
reliable results can be obtained from the TGAS data, corresponds to $\sigma(\varpi)/\varpi<0.35$.

In his table 1 Gontcharov (2016) gave the ranges of distances from the Sun where there is accurate (better than $0.05^m$) photometry for a complete (in these ranges) sample of clump 
giants for different photometric bands. The set of photometric bands that provide a photometric accuracy of at least $0.05^m$ over the entire sky in the entire range of TGAS
magnitudes is very small (the $R_{mode}$ range where the photometry in the band is accurate is given in parentheses): $B_T$ ($<600$ pc) and $V_T$ ($<740$ pc) from
Tycho-2, $G$ (the entire range) from Gaia, $J$ ($>150$ pc), $H$ ($>190$ pc) and $K_s$ ($>200$ pc) from 2MASS (Skrutskie et al. 2006), $W2$ ($>200$ pc), $W3$ (the entire range), 
and $W4$ ($<400$ pc) from WISE (Wright et al. 2010). The following bands and catalogues are unsuitable: $W1$ from WISE does not give accurate photometry for clump giants at
$R_{mode}<500$ pc; $B$, $V$, $g$, $r$, and $i$ from APASS are available for less than half of the clump giants at $R_{mode}<800$ pc; the SDSS, DENIS, and Pan-STARRS photometry 
covers only part of the sky and refers only to faint stars. Thus, only the photometry in the $W3$ and $G$ bands is accurate in the entire region of space within 800 pc of the Sun.

Note that the interstellar extinction in the $G$ band and, in general, in the visual range is much greater than that in $W3$ and, in general, in the infrared (IR) range:
\begin{equation}
\label{ag}
A_G=0.861A_V,
\end{equation}
\begin{equation}
\label{extin}
A_{W3}=0.089A_V,
\end{equation}
according to the extinction law from Weingartner and Draine (2001, hereafter WD2001), or 
\begin{equation}
\label{extinparsec}
A_{W3}=0.002A_V,
\end{equation}
according to the extinction law used in the PARSEC database of theoretical isochrones (Bressan et al. 2012) following the laws from Cardelli et al. (1989)
and O'Donnell (1994). The extinction estimates based on the WD2001 law in the remaining bands with respect to the extinction $A_V$ in the $V$ band were
given by Gontcharov (2016) in his table 1. The lower the extinction, the lesser the contamination of the sample of giants under consideration by reddened
early-type main-sequence (MS) stars and the smaller the change in all of the quantities under consideration due to the stellar reddening. This study is based on
an analysis of the variations of the sum on the left-hand side of the equation (and analogous equations for other bands)
\begin{equation}
\label{basic}
M_{W3}+A_{W3}=W3+5-5\log(R)
\end{equation}
using the variations of the sum on the right-hand side of this equation. Therefore, the analysis is simplified when minimizing the extinction and its spatial
variations as well as the absolute magnitude variations. On the other hand, according to the PARSEC database, the absolute magnitude of a clump giant
depends significantly on the color index (and temperature) for the visual photometric bands and is almost independent of it in the IR range, to be more precise,
at a wavelength longer than 1.5 $\mu$m, i.e., in the $H$, $K_s$, $W1$, $W2$, $W3$, and $W4$ bands. Therefore, the data in these bands are much easier to analyze. Thus, the
main band for this study is $W3$.

As an illustration, Fig. 1 shows a Hertzsprung--Russell (HR) ``$(G-W3)$ -- $M_{W3}$'' diagram in the neighborhoods of the giant clump and branch for stars
with $\sigma(\varpi)/\varpi<0.2$: (a) 9011 Hipparcos stars and (b) 74328 TGAS stars. Since the number of stars with accurate parallaxes is larger by almost an order
of magnitude, TGAS identifies the giant clump much better than Hipparcos does. The stellar positions were not corrected for the reddening $E(G-W3)$ and
extinction $A_{W3}$. This manifests itself for the TGAS stars whose sample extends farther from the Sun than the sample of Hipparcos stars: a tail of reddened
stars is seen to the right of the clump. However, the overwhelming majority of giants are seen to be shifted insignificantly by the reddening and extinction. It is
obvious from the figure that in such a situation the giants are confidently separated from the MS by a set of simple color constrains. The same constraints allow
the nonsingle, peculiar, and misidentified stars to be excluded from consideration. At $R_{mode}>200$ pc, where the $G$, $H$, $K_s$, $W2$, and $W3$ magnitudes
are accurate, we adopted the constraints $G-H>1.5^m$, $G-K_s>1.6^m$, $G-W2>1.6^m$, $G-W3>1.6^m$, $H-K_s>0^m$, 
$H-W2>0^m$, $H-W3>0^m$, $K_s-W2>-0.1^m$, $K_s-W3>-0.1^m$, and $W2-W3<0.25^m$. 
At $R_{mode}<400$ pc, where the $B_T$, $V_T$, $G$, $W3$, and $W4$ magnitudes are accurate, we adopted the constraints 
$B_T-V_T>0.65^m$, $B_T-G>0.9^m$, $B_T-W3>2.5^m$, $B_T-W4>2.5^m$, $V_T-G>0.3^m$, $V_T-W3>2^m$, $V_T-W4>2^m$,
$G-W3>1.6^m$, $G-W4>1.7^m$, $-0.4^m<W3-W4<0.6^m$.
In addition, only the stars with $ccflags=0$ (an image without any influence of the neighboring ones) and $varflg=0-5$ or $varflg=n$
(no noticeable variability) were left in the WISE bands used. Finally, 239794 giants were selected according to the criterion 
$-2.4^m<M_{W3}<-0.8^m$.
The region where the stars were selected is marked in Fig. 1 by the rectangle.

\begin{figure}
\includegraphics{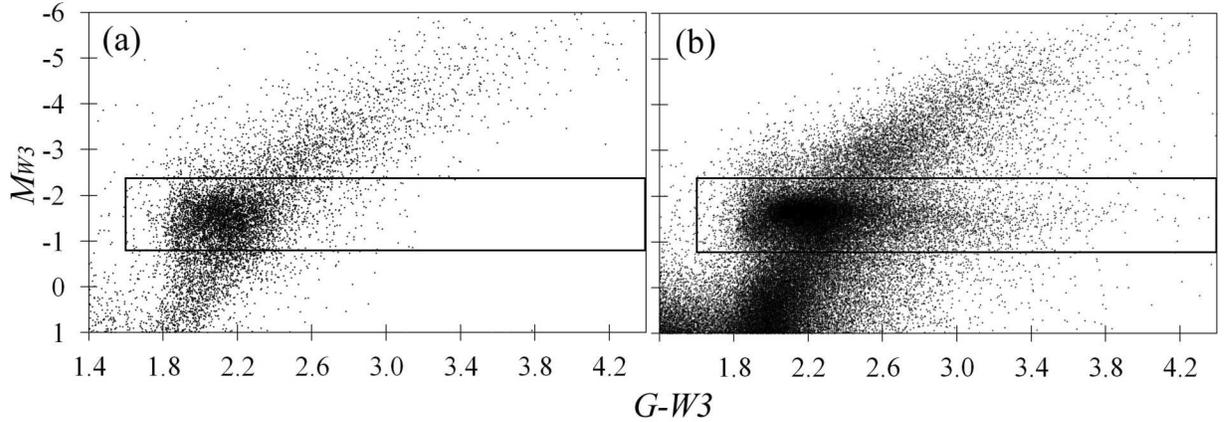}
\caption{Hertzsprung--Russell $(G-W3)$ -- $M_{W3}$ diagram in the neighborhoods of the giant clump and branch for stars with $\sigma(\varpi)/\varpi<0.2$ from Hipparcos (a) 
and TGAS (b). The stellar positions were not corrected for the reddening $E(G-W3)$ and extinction $A_{W3}$. 
The rectangle marks the region where the stars under consideration were selected.
}
\label{hr}
\end{figure}

The giant clump is identified on the HR diagram as a region of enhanced density of stars. Here, branch 
giants are the only significant (up to 20\%, as estimated by Gontcharov (2009)) admixture to the clump
giants. This admixture shifts noticeably the mean and median of the absolute magnitude for the giants
near the clump but does not shift its mode. This is because the branch giants near the maximum of
the distribution of giants on the HR diagram are few in number (less than 10\%) and are distributed quite
uniformly, while the density maximum is determined by the distribution of clump giants. In addition, the
previously adopted constraints on the color indices when they are moderately varied do not shift the mode
of the absolute magnitude for all of the remaining stars in the sample either, in contrast to the mean and
median. Therefore, it is the mode that was chosen in this study as an estimate of the typical absolute magnitude
for clump giants. This allowed us to restrict ourselves to the mentioned rough selection of stars
and to avoid a more refined selection with inevitable uncertainties and biases.

To calculate $mode(M_{W3})$ corresponding to the maximum of the distribution of giants, we counted
the number of giants in $\Delta M_{W3}=0.01^m$ cells. To analyze the dependence of $mode(M_{W3})$ on a particular
parameter, we ranked the stars by an increase in this parameter and calculated $mode(M_{W3})$ similarly
to a moving average: for the subsample of stars with ordinal numbers from 1 to $i$ (with the minimum value
of the parameter), then from 2 to $i+1$, from 3 to $i+2$, etc. As a result, we obtained a set of $mode(M_{W3})$ that
can be associated with the set of similarly averaged values of the parameter under consideration. The
window $i$ was chosen so that the number of stars in each subsample under consideration was sufficient
to determine the maximum of their distribution but so that the number of subsamples was large enough
to determine the dependence of $mode(M_{W3})$ on the parameter. For example, at the previously mentioned
reasonable sample constraint $\sigma(\varpi)/\varpi<0.35$, 97468 giants
\footnote{These should be considered in future studies as a sample of clump giants with the most accurate parallaxes, although in
this study the $\sigma(\varpi)/\varpi$ constraint is not applied everywhere.}
remain in the sample. Given that the uncertainty of the photometry for the stars under
consideration is much smaller than $\sigma(\varpi)/\varpi$, the error in $M_{W3}$ for one star does not exceed 
$0.35\times2.17=0.76^m$ (Parenago 1954, p. 44). Adopting $i=6000$ and taking into account the scatter of giants in color
and the influence of branch giants, we can count on the determination of $mode(M_{W3})$ with an accuracy
of a few hundredths of a magnitude. At $i<6000$ the results have a low accuracy. Therefore, this method is
efficient only for large samples, at least several tens of thousands of stars. It could not be applied to the data
from the Hipparcos catalogue containing no more than 10 000 giants with accurate parallaxes.

\begin{figure}
\includegraphics{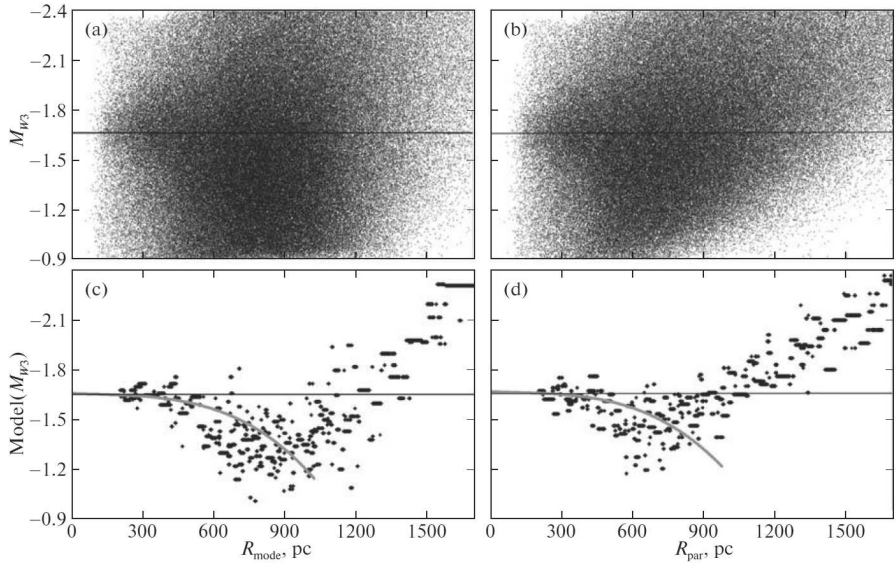}
\caption{Positions of all the giants under consideration on the $R_{mode}$ -- $M_{W3}$ (a) and $R_{par}$ -- $M_{W3}$ (b) diagrams. The dependences
of $mode(M_{W3})$ on $R_{mode}$ (c) and $R_{par}$ (d) for all of the giants under consideration. When calculating $M_{W3}$, we used $R_{mode}$ or
$R_{par}$, respectively. The horizontal lines mark $M_{W3}=-1.66^m$. The thick gray curves on panels (c) and (d) indicate the influence
of the error from Eq. (7). For comparison with the HR diagram, the vertical axis in this and succeeding figures has the reverse order of values.
}
\label{rmodeall}
\end{figure}

\section*{RESULTS}

Figure 2 shows the positions of all the giants under consideration (without any constraint on $\sigma(\varpi)/\varpi$) on
the $R_{mode}$ -- $M_{W3}$ (a) and $R_{par}$ -- $M_{W3}$ (b) diagrams as well as the dependences of $mode(M_{W3})$
on $R_{mode}$ (c) and $R_{par}$ (d) for all giants. We calculated $M_{W3}$ using $R_{mode}$ or $R_{par}$, respectively. 
The horizontal line marks $M_{W3}=-1.66^m$. For comparison with the HR diagram, the vertical axis in this and succeeding
figures has the reverse order of values.

In Figs. 2a and 2b the giant clump manifests itself precisely as a region of enhanced density of stars.
Many of the stars with very large $R_{par}$ are seen to have much smaller $R_{mode}$. This was described and justified
by Astraatmadja and Bailer-Jones (2016). It can also be seen that, despite the differences in
distances, Figs. 2c and 2d show approximately the same dependence of $mode(M_{W3})$ on the distance,
\begin{equation}
\label{esti}
mode(M_{W3})=-1.66^m\pm0.02^m
\end{equation}
near the Sun, $mode(M_{W3})$ then decreases in absolute value with increasing distance and increases in absolute
value from some critical distance. This distance is $R_{mode}\approx800$ and $R_{par}\approx700$ pc. A comparison of
$M_{W3}$ and $mode(M_{W3})$ shows that the giant clump is more saturated with stars at distances smaller than
the critical one, i.e., the sample is more complete than that at distances larger than the critical one. At
large distances we deal with the branch giants in its upper part instead of the clump giants. They have
a much smaller absolute magnitude and, hence, are visible at greater distances. This can be seen from a
comparison of Figs. 2a and 2b with Fig. 1b. When using $R_{par}$, many of the clump giants with large $R_{par}$
are scattered among the branch giants. When using $R_{mode}$, they ``move'' closer to the Sun and make the
sample of clump giants more complete at intermediate distances $700<R_{mode}<800$ pc. This explains
the above difference in critical distances $R_{mode}\approx800$ and $R_{par}\approx700$ pc. 
Thus, as expected, (1) $R_{mode}$ is closer to the true one than $R_{par}$, and (2) a complete (to some degree) sample of clump 
giants within 800 pc of the Sun is obtained when using $R_{mode}$ and $W3$ photometry.

However, the significant systematic change of $mode(M_{W3})$ at distances smaller than the critical one
(i.e., at $400<R_{mode}<800$ pc and $400<R_{par}<700$ pc), where the mentioned selection effect should
not be present, is much more interesting. Here, some effect that needs to be explained counteracts the selection effect.

Figure 3 shows the positions of all the giants under consideration on the $W3$ -- $M_{W3}$ (a) and $G$ -- $M_G$ (b)
diagrams and the dependences of $mode(M_{W3})$ on $W3$ (c) and $mode(M_G)$ on $G$ (d) for them. The vertical
axes have different scales. The horizontal lines mark $M_{W3}=-1.66^m$ and $M_G=0.44^m$. The brightest
stars are grouped around these values. As in Figs. 2a and 2b, in Figs. 3a and 3b the giant clump manifests
itself as a region of enhanced density of stars. We see that the bulk of the stars shift downward with increasing $W3$
and $G$. This causes a systematic change of $mode(M_{W3})$ in Figs. 3c and 3d. This dependence
is also retained when the sample is constrained in $\sigma(\varpi)/\varpi$. In contrast to Figs. 2a and 2b, in Figs. 3a
and 3b the branch giants do not dominate at any $W3$ and $G$. Consequently, the observed systematic trend
is explained by selection only to a small extent and is mainly caused by a different effect.

Thus, the main question of this study reflected in Figs. 2 and 3 is the following: Why does $mode(M_{W3})$
change systematically with $R_{mode}$ or $W3$ at $400<R_{mode}<800$ pc and $W3>6.5^m$? This question
is very important, given that clump giants are actively used as standard candles, while the difference
between $mode(M_{W3})\approx-1.7^m$ at $W3\approx6^m$ and $mode(M_{W3})\approx-1.0^m$ at $W3\approx9.2^m$ 
gives an unacceptably large relative error of 32\% when calculating the distances from the absolute magnitudes.

Let us test whether the observed effect is caused by interstellar extinction variations. According to
Eq. (4), we actually observe a systematic change of $mode(M_{W3}+A_{W3})$ in the range at least from $-1.7^m$ to $-1.0^m$, 
i.e., by $0.7^m$, and a change of $mode(M_G+A_G)$ approximately by $0.8^m$. In the space under
consideration a systematic change of $mode(A_G)$ by $0.8^m$ is unlikely, but admissible. However,
$mode(A_{W3})$ can then change by no more than $0.08^m$ according to Eqs. (1) and (2) and only by
$0.002^m$ according to Eq. (3). The difference between the extinction laws from Eqs. (2) and (3) also allows
the systematic variations of $mode(A_{W3})$ due to the possible spatial variations of the extinction law, which
were pointed out, for example, by Gontcharov (2016), to be roughly estimated: less than $0.07A_G$. Thus, all
of the mentioned variations in extinction and its law can explain no more than $0.08^m$ from the total $0.7^m$
change of $mode(M_{W3}+A_{W3})$ under consideration. The rest needs to be explained by a change in $M_{W3}$ or
the quantities on the right-hand side of Eq. (4).

\begin{figure}
\includegraphics{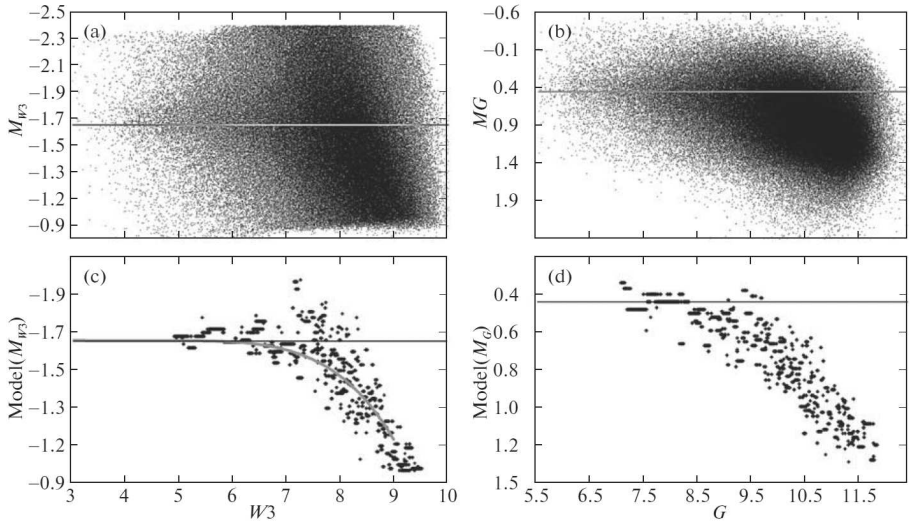}
\caption{Positions of all the giants under consideration on the $W3$ -- $M_{W3}$ (a) and $G$ -- $M_G$ (b) diagrams. The dependences
of $mode(M_{W3})$ on $W3$ (c) and $mode(M_G)$ on $G$ (d) for all of the giants under consideration. The horizontal lines mark
$M_{W3}=-1.66^m$ and $M_G=0.44^m$. The thick gray curve on panel (c) indicates the influence of the error from Eq. (7). The vertical axes have different scales.
}
\label{w3mw3gmg}
\end{figure}

The change by $0.7^m$ under consideration is much larger than the natural scatter of $M_{W3}$ for the clump
giants caused by variations in their metallicity and age in the space under consideration. This can be
seen from Fig. 4, which shows the isochrones of stars with ages of 2 (black solid curves), 5 (gray curves),
and 10 (black dotted curves) Gyr and metallicities 
\footnote{To avoid confusion, everywhere below the metallicity is designated as $\mathbf Z$, 
while one of the Galactic coordinates is designated as $Z$.}
$\mathbf Z=0.006$, 0.009, 0.012, 0.015, and 0.018 (five curves from left to right for each age, respectively) in the
neighborhoods of the giant clump, i.e., at a comparatively long evolutionary stage with helium burning
in the core and an inert hydrogen envelope (the subsequent beginning of nuclear reactions in the envelope
rapidly moves the giant away from the clump on the asymptotic branch to a different region of the HR diagram). 
The horizontal line marks $M_{W3}=-1.66^m$. Gontcharov (2016) showed that within the
kiloparsec nearest to the Sun the mean metallicity is $\mathbf Z>0.006$ everywhere. On the other hand, the
solar metallicity estimate, which has changed in recent years from $\mathbf Z=0.019$ (Marigo et al. 2008) to
$\mathbf Z=0.015$ (Bressan et al. 2012), also fits into the $\mathbf Z$ range presented in the figure. Thus, the natural
variations of $mode(M_{W3})$ due to the variations in the mean age and mean metallicity of the sample within
800 pc of the Sun must fit into the range between the isochrones shown in Fig. 4, i.e., into the range
$-1.86^m<M_{W3}<-1.48^m$, or 
\begin{equation}
\label{izomw3}
M_{W3}=-1.67^m\pm0.19^m.
\end{equation}
Bovy et al. (2014) obtained an even narrower range of variations, $M_{K_s}=-1.65^m\pm0.025^m$, by analyzing
it using the PARSEC database as a function of the properties of the stellar population in the solar
neighborhoods. The theoretical scatters of $M_{K_s}$ and $M_{W3}$ due to the age and metallicity variations
must be approximately identical. Therefore, as a result, the theoretical $M_{W3}$ estimates in the range
$-1.6^m\div-1.7^m$ seem more plausible than those in the range $-1.4^m\div-1.6^m$.

Thus, taking into account the theoretical estimate (6) for $M_{W3}$, we must assume that from
the empirical estimates based on Figs. 2 and 3, $mode(M_{W3})=-1.66^m$ at $R_{mode}<400$ pc and $W3<6.5^m$ or 
$mode(M_{W3})\approx-1.35^m$ at $R_{mode}\approx800$ pc and $W3\approx8.5^m$, the former is more likely closer to the
true one, while the latter is plagued with a systematic error that is not related to the natural variations of
$mode(M_{W3})$.

Thus, having rejected the implausibly large changes of $M_{W3}$ and $A_{W3}$ in the space under consideration,
we must attribute the changes of the right-hand side of Eq. (4) to the change in distance and, hence, to the
systematic errors of the TGAS parallax.

\begin{figure}
\includegraphics{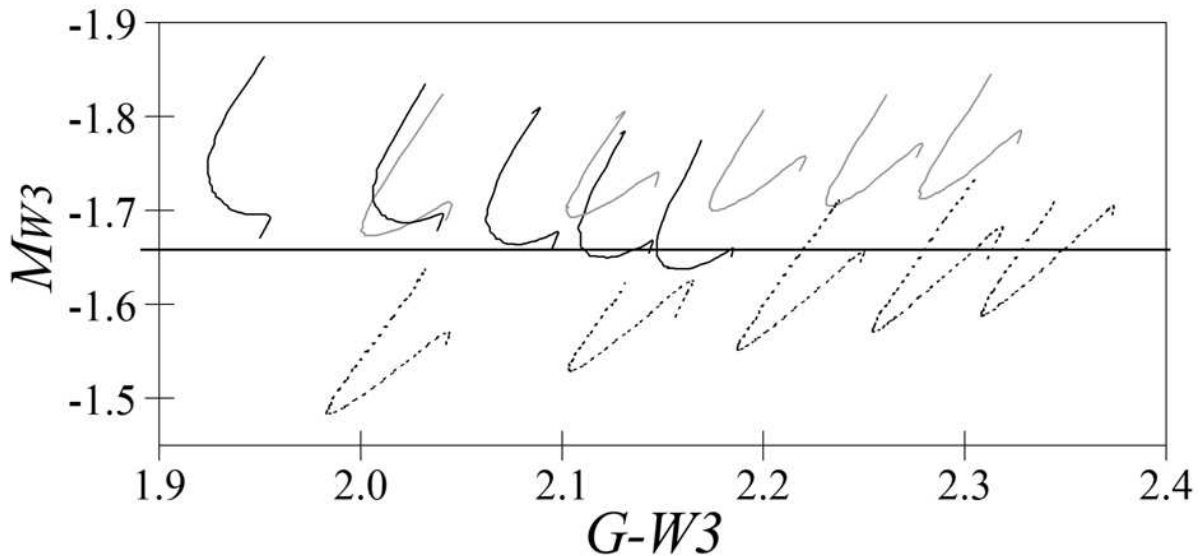}
\caption{Isochrones of stars with ages of 2 (black solid curves), 5 (gray curves), and 10 (black dotted curves) Gyr and metallicities
$\mathbf Z=0.006$, 0.009, 0.012, 0.015, and 0.018 (five curves from left to right for each age, respectively) in the neighborhoods of the
giant clump. The horizontal line marks $M_{W3}=-1.66^m$.
}
\label{izo}
\end{figure}

\section*{USING THE PHOTOMETRIC DISTANCES}

Let us make sure that the distances for the same stars but from a different source do not show such
large variations of the right-hand side of Eq. (4). The method of this study is inapplicable to the Hipparcos
parallaxes, because this catalogue contains only 9344 giants with $\sigma(\varpi)/\varpi<0.35$, and almost all of them
are closer than 400 pc to the Sun, where the sought-for effect barely manifests itself.

Another source of independent distances for clump giants is the catalogue by Gontcharov (2008). It
contains 97348 suspected clump giants selected from the Tycho-2 catalogue using their photometry and
reduced proper motions. Their photometric distances $R_{g2008}$ were calculated with a relative error of 0.25 by
taking into account the extinction estimate based on the stellar spectral energy distribution and assuming
$M_{K_s}=-1.52^m$. TGAS contains 87993 of these stars. Being samples from the same original Tycho-2 catalogue,
the sample from Gontcharov (2008) and the TGAS sample obtained in this study occupy approximately
the same space, predominantly the kiloparsec nearest to the Sun, including the range of distances
where the effect being investigated is observed. For 62885 stars with $\sigma(\varpi)/\varpi<0.35$ it makes sense
to compare $R_{g2008}$ and $R_{mode}$: they agree well. At the same time, the positions of these stars on the
HR diagram using the TGAS parallaxes shows that most (78\%) of them actually belong to the giant
clump. The admixture of branch giants and MS stars is about 20\% and less than 0.5\%, respectively.

85620 stars from the sample by Gontcharov (2008) correspond to the photometric constraints applied
above in this study. For them Fig. 5 shows the dependences of $mode(M_{W3})$ calculated using $R_{g2008}$
(black dots) and $R_{mode}$ (gray dots) on the distances themselves (a) and $W3$ (b). The variations of
$mode(M_{W3})$ with $R_{g2008}$ are so small that the row of the black dots in the figure merged into a solid curve.
The solid and dashed horizontal lines mark $M_{W3}=-1.66^m$ and $-1.53^m$, respectively. As previously, the
first value is seen to describe well $mode(M_{W3})$ for the nearest and brightest clump giants using $R_{mode}$.
The second value is given for the same stars but using $R_{g2008}$. It roughly corresponds to the zero
point $M_{K_s}=-1.52^m$ adopted by Gontcharov (2008). Thus, the difference between the zero points is
$\Delta M_{W3}=-1.66-(-1.53)=-0.13^m$. The values of $mode(M_{W3})$ calculated from $R_{g2008}$ (black dots)
were shifted vertically by this value in Figs. 5c and 5d.

The black dots show a monotonic increase in $mode(M_{W3})$ with $R_{g2008}$ and $W3$ caused by the
growing selection in favor of the previously noted admixture of branch giants in the sample by Gontcharov
(2008). The gray dots everywhere follow the falls and rises from Figs. 2c and 3c. Bringing the zero points in
Figs. 5c and 5d into coincidence shows the ranges of $R_{mode}$ and $W3$ in which the gray dots systematically
deviate from the black ones, i.e., the ranges where the systematic error of the TGAS parallaxes primarily
manifests itself. Thus, once the zero points have been brought into coincidence, we see that the results from
Gontcharov (2008) are affected only by selection, while the results from TGAS are affected by two
effects: selection and the parallax error.

So, the dependences for the TGAS data found in this study are retained when the sample composition
is slightly changed, are unique only to these data, and do not manifest themselves when using a different set of distances.

\begin{figure}
\includegraphics{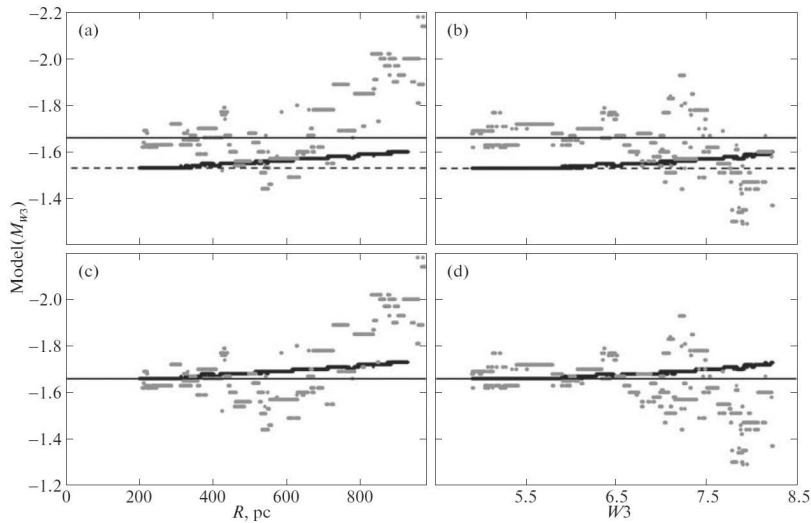}
\caption{Dependences of $mode(M_{W3})$ calculated using $R_{g2008}$ (black dots) and $R_{mode}$ (gray dots) on the distances themselves
(a) and $W3$ (b). The solid and dashed horizontal lines mark $M_{W3}=-1.66^m$ and $-1.53^m$, respectively; (c) and (d)
the same but the zero point of $R_{g2008}$ was reduced to $M_{W3}=-1.66^m$.
}
\label{gaiarcg}
\end{figure}

\section*{THE SYSTEMATIC PARALLAX ERROR}

The dependences of $mode(M_{W3})$ on $W3$ and $G$ from Figs. 3c and 3d resemble the dependence
of $mode(M_{W3})$ on $R_{mode}$ from Fig. 2c at $R_{mode}<800$ pc. This is because there is a correlation between
the distance and magnitude. The error $\sigma(\varpi)/\varpi$, the de-reddened color indices (due to the correlation of
the distance, reddening, and interstellar extinction), the absolute values of the proper motion components
$|\mu_{\alpha}|$ and $|\mu_{\delta}|$, the total proper motion $(\mu_{\alpha}^2+\mu_{\delta}^2)^{1/2}$,
their relative errors, and the Galactic coordinates $X$, $Y$, and $Z$ also clearly correlate with the distance
and magnitude. The revealed dependences of $mode(M_{W3})$ on these parameters are completely
explained by these correlations.

Excluding the stars with large $\sigma(\varpi)/\varpi$ from the sample is equivalent to excluding the stars with large
$R_{mode}$ due to their correlation. Therefore, to avoid the influence of sample incompleteness at $R_{mode}>800$ pc 
on the results, here and below we consider only 97468 giants with $\sigma(\varpi)/\varpi<0.35$.

As an example, Figs. 6a--6c show the dependences of $mode(M_{W3})$ on $X$, $Y$, and $Z$, respectively.
The drop in $mode(M_{W3})$ at large $R_{mode}$ manifests itself as a drop in $mode(M_{W3})$ at large $|X|$, $|Y|$, and $|Z|$. 
In addition, from eight sectors of the Galaxy (four quadrants $\times$ two hemispheres) the sample of
giants under consideration extends farthest in the sector with negative $X$, $Y$, and $Z$ (the third quadrant
below the Galactic midplane, the Orion--Eridanus region, approximately between the south Galactic and
ecliptic poles). Here, the drop in $mode(M_{W3})$ at large $R_{mode}$ also manifests itself as a drop in $mode(M_{W3})$
at negative $X$, $Y$, $Z$ and minimum $b$ and $\beta$. This can be seen from Figs. 6d--6h, where the dependences of
$mode(M_{W3})$ on right ascension $\alpha$, ecliptic latitude $\beta$ and longitude $\lambda$, and Galactic latitude $b$ 
and longitude $l$, respectively, are shown.

\begin{figure}
\includegraphics{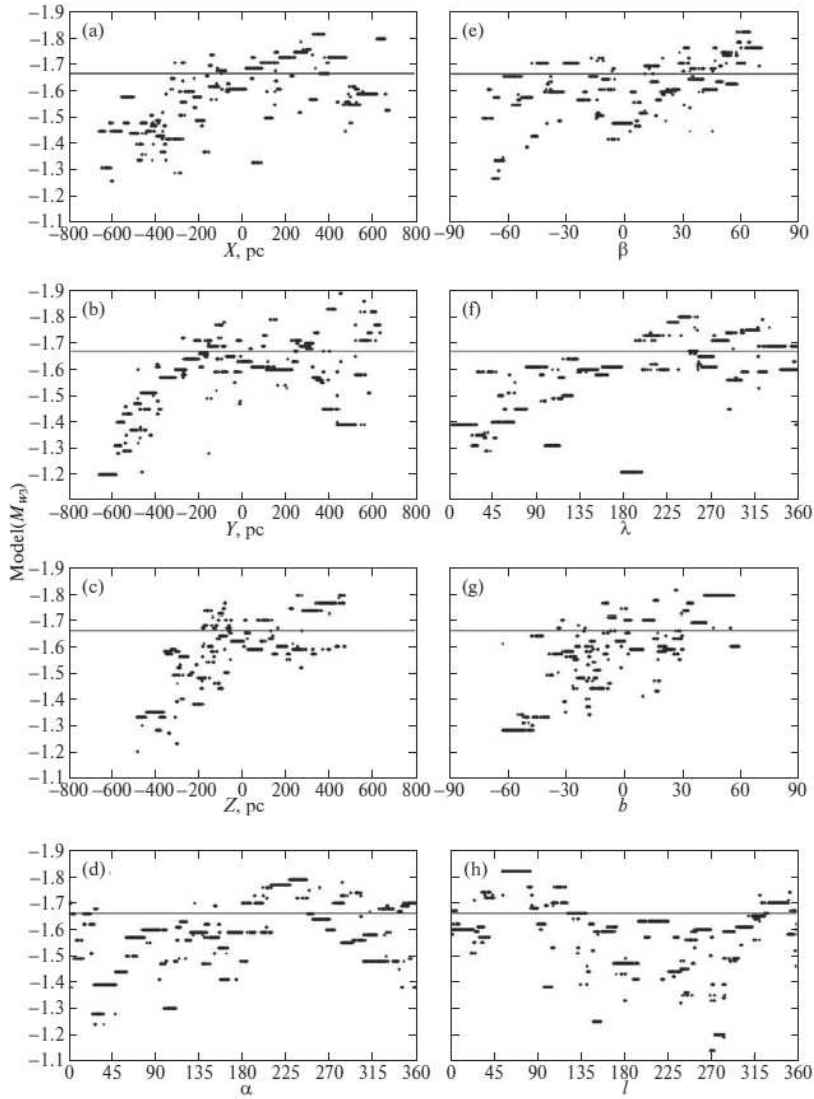}
\caption{Dependences of $mode(M_{W3})$ on $X$ (a), $Y$ (b), $Z$ (c), $\alpha$ (d), $\beta$ (e), $\lambda$ (f), $b$ (g), and $l$ (h) for giants with $\sigma(\varpi)/\varpi<0.35$.
The horizontal line marks $M_{W3}=-1.66^m$.
}
\label{depend}
\end{figure}

\begin{figure}
\includegraphics{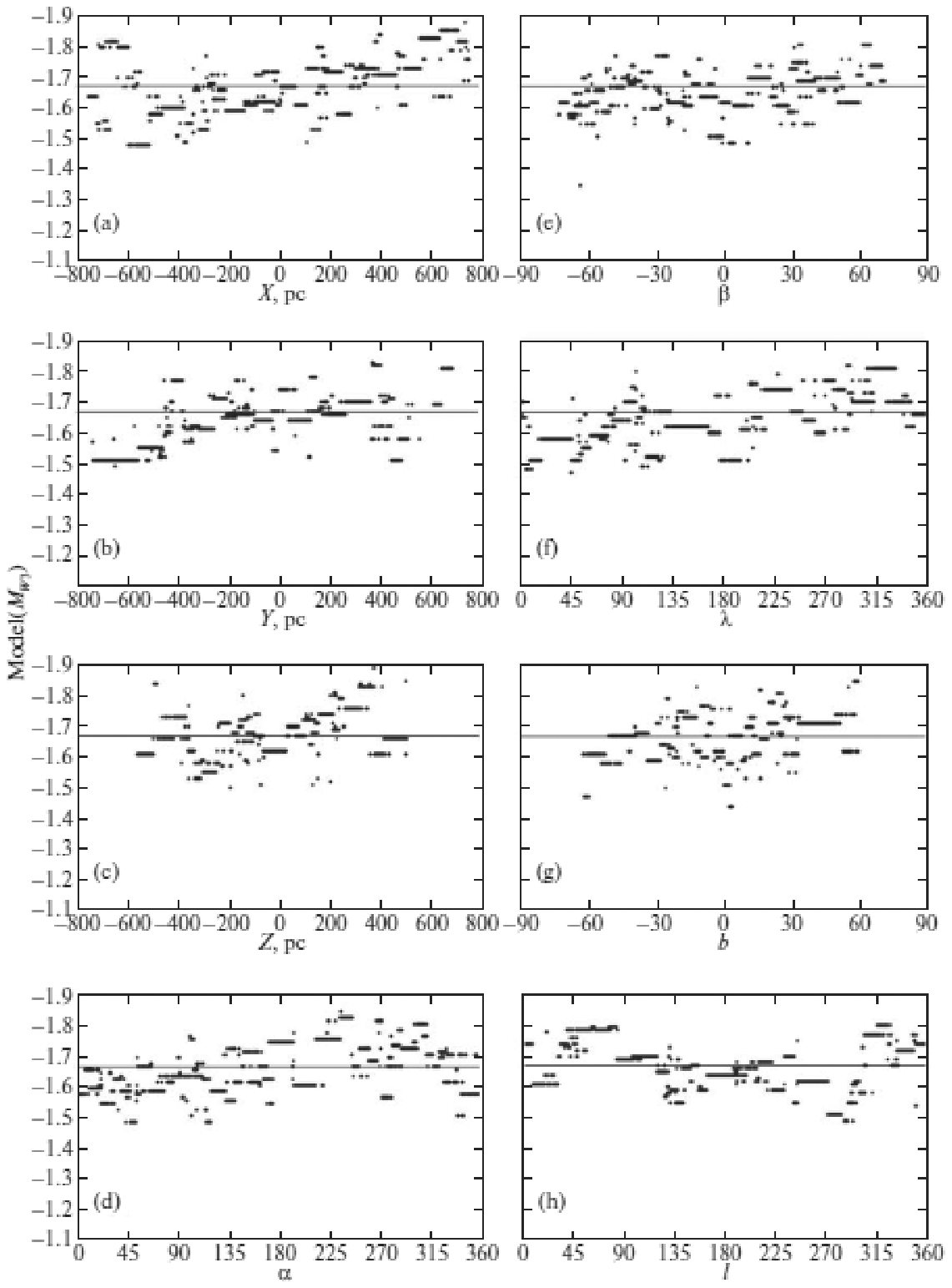}
\caption{Same as Fig. 6 after the correction of the parallaxes for the error (7).
}
\label{dependcor}
\end{figure}

In some regions of the sky, according to Fig. 7 from Lindegren et al. (2016), $R_{mode}$ correlates with
$\alpha$, $\delta$, and $\mu_{\alpha}$ via $\varpi$. 
As a result, the dependences of $mode(M_{W3})$ on these quantities arise. This is particularly clearly seen in the sharp decrease in
$mode(M_{W3})$ at $-30^{\circ}<\alpha<60^{\circ}$ in Fig. 6d. The dependence of $mode(M_{W3})$ on $\delta$ is not so clear and,
therefore, is not shown in the figure. 

The dependence of $mode(M_{W3})$ on $W3$ seen in Fig. 3c can be explained by a parallax error 
proportional to the square of the measured distance:
\begin{equation}
\label{corr}
\Delta\varpi=0.18R_{mode}^2,
\end{equation}
where $R_{mode}$ is in kpc and $\varpi$ is in mas. The coefficient of this dependence was found by the least-squares method. The influence of this error on the
dependences of $mode(M_{W3})$ on $R_{mode}$ and $W3$ is indicated in Figs. 2 and 3 by the thick gray curves, respectively. Correcting the parallaxes for the error
using Eq. (7) leads to a significant decrease in the variations of $mode(M_{W3})$ in the dependences on all parameters shown in Fig. 6. The same dependences
after the correction are shown in Fig. 7.

The dependences on the parameters are removed slightly more poorly if the dependence of $mode(M_{W3})$ on $W3$ is explained by an error proportional to the
magnitude:
\begin{equation}
\label{corw3}
\Delta\varpi=0.09W3-0.585
\end{equation}
for stars with $W3>6.5^m$ (the coefficients were found by the least-squares method). The jumps in $mode(M_{W3})$ at $W3\approx7.3^m$ and $mode(M_G)$ at
$G\approx9.3^m$ are seen in Figs. 3c and 3d. This may be how different dependences manifest themselves for different magnitude ranges.

In any case, this error is very small compared to the accuracy of the TGAS parallaxes declared by the authors. At the true distance of 100 pc we have the true parallax and its error from 
Eq. (7)
$\varpi=10.000\pm0.002$, at 500 pc $\varpi=2.000\pm0.043$, at 800 pc $\varpi=1.250\pm0.099$, at 1000 pc $\varpi=1.000\pm0.139$, at 1500 pc $\varpi=0.667\pm0.226$ mas.
Because of this error, the true parallax is always smaller, while the distance is larger than the measured one.

Let us assess how the parallax error found agrees with the tests of the TGAS parallaxes mentioned in the Introduction. According to Eq. (7), the mean presumed error of the TGAS 
parallaxes for giants with $\sigma(\varpi)/\varpi<0.35$, i.e., the ``TGAS minus clump giants'' difference, is $\Delta\varpi=0.06$ mas. For the
Hipparcos stars used by Lindegren et al. (2016) it is $\Delta\varpi=0.037$ mas for the entire sky, 0.036 and 0.038 mas northward and southward of the ecliptic,
respectively. Given the ``TGAS minus Hipparcos'' difference $\Delta\varpi=-0.089$ mas found by Lindegren et al. (2016), we obtain the ``clump giants minus Hipparcos'' 
difference $\Delta\varpi=-0.089-0.037=-0.126$ mas, while taking their values northward and southward of the ecliptic, we obtain $-0.166$ and $-0.091$ mas, respectively. 
For the ``clump giants minus Hipparcos'' differences we found the same trends as a function of the color index $V-I$ as those found by Lindegren et al. (2016) for the 
``TGAS minus Hipparcos'' difference: a rise in $\Delta\varpi=0.09(V-I)$ and a drop in $\Delta\varpi=-0.11(V-I)$ northward and southward of the ecliptic, respectively. This also
manifests itself for a different color: $\Delta\varpi=0.045(G-W3)$ northward and $\Delta\varpi=-0.059(G-W3)$ southward.

It may well be that the negative ``clump giants minus Hipparcos'' and ``TGAS minus Hipparcos'' differences $\Delta\varpi$ and the negative difference found by
Jao et al. (2016) also for the Hipparcos stars are manifestations of the Hipparcos parallax error. To bring the ``clump giants minus Hipparcos'' difference
$\Delta\varpi$ closer to zero, we would have to reverse the sign of (7), but the dependences of $mode(M_{W3})$ on $R_{mode}$, $R_{par}$, $W3$, and $G$ shown in Figs. 2 and 3 for
stars with $\sigma(\varpi)/\varpi<0.35$ (i.e., in fact, for $R_{mode}<800$ pc and $R_{par}<700$ pc) would then strengthen and remain without an explanation. However, the
trend at $R_{mode}>800$ pc and $R_{par}>700$ pc seen in Fig. 2 would then vanish in this case, but, as has been discussed previously, this trend is caused not by the
parallax error but by the selection in favor of higher luminosity stars. Stassun and Torres (2016) pointed out that the difference $\sigma(\varpi)/\varpi$ found by them approaches
zero for distant stars with $\varpi<1$ mas. This may imply that the distances of the eclipsing binary stars used by them were determined without allowance for
this selection and, hence, contain an error increasing with distance. This is indirectly confirmed by the result from De Ridder et al. (2016) mentioned in the
Introduction: there is agreement of the TGAS parallaxes with the asteroseismic ones for nearby stars and disagreement for distant ones. The remaining
tests with quasars and Cepheids mentioned in the Introduction showed no significant difference between their parallaxes and the TGAS parallaxes, on average,
over the entire sky. 

The dependences found in this study were successfully removed when (7) was subtracted from the parallaxes. This suggests that the TGAS parallaxes,
at least for the giants within 800 pc of the Sun, are free from any systematic errors exceeding some limit. It can be determined if we take into account the largest
systematic variations of $mode(M_{W3})$ remained after the correction for the error (7) and seen in Fig. 7. These variations in combination with the removed
ones correspond to an error in $\varpi$ of no more than 0.2 mas within 800 pc of the Sun.

\begin{figure}
\includegraphics{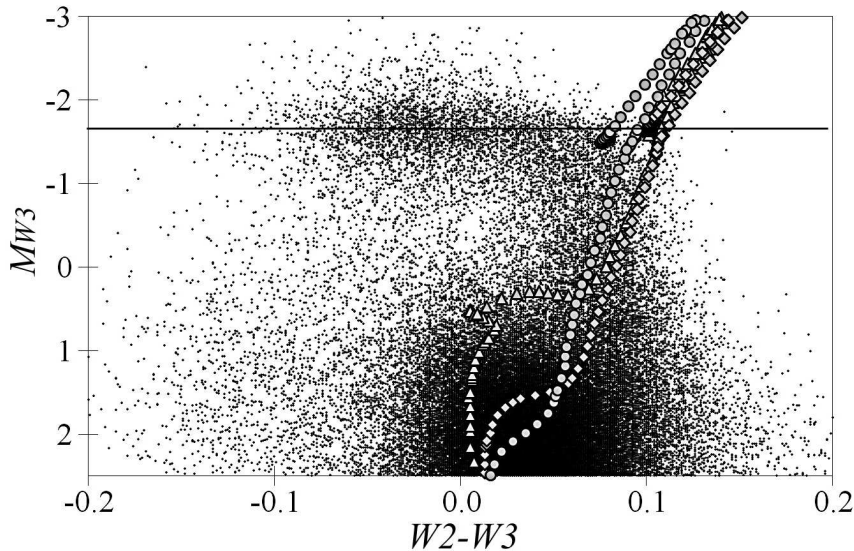}
\caption{Isochrones for $\mathbf Z=0.018$ and an age of 2 Gyr (gray triangles), $\mathbf Z=0.012$ and 5 Gyr (gray diamonds), and $\mathbf Z=0.006$
and 10 Gyr (gray circles) against the background of an HR ``$(W2-W3) - M_{W3}$'' diagram for all TGAS stars with $|b|>40^{\circ}$,
accurate $W2$ and $W3$ photometry, and $\sigma(\varpi)/\varpi<0.2$. The horizontal line marks $M_{W3}=-1.66^m$.
}
\label{w2w3}
\end{figure}

\section*{THE ABSOLUTE MAGNITUDES AND COLORS OF CLUMP GIANTS}

In this study, similarly to the $W3$ band, we also analyzed the other mentioned bands. As a result, the modes of the absolute magnitudes and dereddened
colors in some bands can be determined for clump giants near the Sun. In the previous studies of different authors such estimates were distorted by the uncertainties
in the extinction and reddening estimates (especially in the visual range), the systematic errors of the distances used (for example, by the Lutz-Kelker and Malmquist biases), 
the uncertainty due to a significant dependence of the absolute magnitude on color for the visual bands, and the selection in favor of faint and distant stars due to the images
of bright and nearby ones being overexposed in the IR range. Therefore, the empirical and theoretical estimates have often disagreed until now.

For example, based on the 2MASS photometry and Hipparcos parallaxes, Groenewegen (2008) obtained an empirical estimate of the mean absolute
magnitude for clump giants, $M_{K_s}=-1.54\pm0.04$. Having supplemented the 2MASS data with their own observations of bright clump giants, Laney
et al. (2012) obtained better estimates: $M_J=-0.984\pm0.014$, $M_H=-1.490\pm0.015$, and $M_{K_s}=-1.613\pm0.015$. The discrepancy between the latter value
and their own theoretical estimate $M_{K_s}=-1.65^m\pm0.025^m$ was noted by Bovy et al. (2014).

An example of a great discrepancy between the estimates is shown in Fig. 8. It shows the theoretical isochrones for $\mathbf Z=0.018$ and an age of 2 Gyr (gray triangles), 
$\mathbf Z=0.012$ and 5 Gyr (gray diamonds), and $\mathbf Z=0.006$ and 10 Gyr (gray circles) against the background of an HR $(W2-W3)$ -- $M_{W3}$ diagram
for all (both giant and MS) TGAS stars at high latitudes ($|b|>40^{\circ}$) with accurate (better than $0.05^m$) $W2$ and $W3$ photometry and $\sigma(\varpi)/\varpi<0.2$. 
We see a good correspondence of the isochrones to the actual distribution of MS stars (bottom) but a very poor correspondence for giants (top). The giant clump is
identified as an isolated group of stars along the horizontal line marking $M_{W3}=-1.66^m$. We see that the theory predicts their color poorly but reproduces well
$mode(M_{W3})\approx-1.66^m$. Consequently, the theory is valid for $W3$ but erroneous for $W2$.

The results from Yaz G\"okce et al. (2013) are the most comparable with the results of this study for the key $W3$ band: these authors considered
$mode(M_{W1})$ and $mode(M_{W3})$. They found a strong dependence of these quantities on the sample constraint in $\sigma(\varpi)/\varpi$: 
from $mode(M_{W1})=-1.576^m\pm0.024^m$ and $mode(M_{W3})=-1.552^m\pm0.020^m$ at $\sigma(\varpi)/\varpi<0.05$ to 
           $mode(M_{W1})=-1.635^m\pm0.026^m$ and $mode(M_{W3})=-1.606^m\pm0.024^m$ at $\sigma(\varpi)/\varpi<0.15$. 
As a result, they adopted the values at $\sigma(\varpi)/\varpi<0.15$ as the best ones. However, the constraint $\sigma(\varpi)/\varpi<0.15$ does not guarantee the sample completeness
and the absence of biases either. Therefore, the errors of these results, as half of the difference of two values, i.e., $\sigma(mode(M_{W1}))=0.03^m$ and 
$\sigma(mode(M_{W3}))=0.03^m$, should be estimated more conservatively. These uncertainties should be increased approximately to $0.06^m$ if we take into
account the large discrepancy found by the authors between their empirical estimate of $mode(M_{W3})=-1.606^m\pm0.024^m$ and the estimate of
$mode(M_{W3})=-1.676^m\pm0.028^m$ obtained by them based on the photometry in other IR bands with allowance made for the typical spectral energy distribution of a clump
giant. They also estimated the absolute magnitudes of clump giants in other bands: 
$mode(M_J)=-0.970^m\pm0.016^m$ 
$mode(M_H)=-1.462^m\pm0.014^m$ 
$mode(M_{K_s})=-1.595^m\pm0.025^m$.

Thus, a noticeable discrepancy between the empirical and theoretical estimates grouping around  $M_{W3}\approx-1.61^m$ and $M_{W3}\approx-1.67^m$, respectively,
was found in previous studies. This study apparently removes this contradiction primarily owing to the accurate Gaia parallaxes. For the stars with the most
accurate data we obtained an empirical estimate of (5) close to the theoretical estimate of (6) and showed the sources of the previous discrepancy: the systematic
errors and selection.

Analysis of all accurate photometric data for the clump giants nearest to the Sun under the assumption of negligible extinction and reddening allowed
mutually consistent estimates to be obtained for the dereddened IR colors and absolute magnitudes presented in the table. The absence of photometry for
many of the brightest stars and the presence of some reddening and extinction even within 200 pc of the Sun are primarily responsible for the uncertainties in
these quantities, which are also presented in the table.

\section*{CONCLUSIONS}

This study is the first mutual test for the Gaia DR1 TGAS parallaxes and characteristics of clump giants. The study showed that, in accordance with the theory,
the giant clump actually creates an enhanced density of stars on the HR diagram with fairly small admixtures and a color independence of the IR absolute
magnitudes. Thus, clump giants are convenient as standard candles. However, the systematic dependences of the absolute magnitudes of clump giants on
various parameters found in the study forces them to be carefully taken into account in future. We showed that removing the key dependence on the distance or
magnitude not only reduced considerably the remaining dependences but also eliminated all discrepancies between the theoretical and empirical estimates of the
characteristics of these stars and demonstrated the universality of these characteristics, at least within 800 pc of the Sun. Using the Gaia parallaxes, we
have obtained accurate estimates of the IR absolute magnitudes for clump giants near the Sun for the first time: 
$mode(M_H)=-1.49^m\pm0.04^m$,
$mode(M_{K_s})=-1.63^m\pm0.03^m$,
$mode(M_{W1})=-1.67^m\pm0.05^m$
$mode(M_{W2})=-1.67^m\pm0.05^m$,
$mode(M_{W3})=-1.66^m\pm0.02^m$,
$mode(M_{W4})=-1.73^m\pm0.03^m$,
and the corresponding de-reddened colors (given in the table). Clump giants turned out to be a convenient tool for testing the systematic parallax errors. We
showed a high accuracy of the Gaia DR1 TGAS parallaxes: within 800 pc of the Sun their systematic errors apparently do not exceed 0.2 mas and are well
described by simple dependences on the distance squared or magnitude. The TGAS catalogue is a good data source for further studies of giants within the nearest kiloparsec.

\textsf{
\begin{table*}
\def\baselinestretch{1}\normalsize\normalsize
\caption[]{Empirical estimates of the mode of the de-reddened color and absolute magnitude with an indication of its mean
error for clump giants near the Sun.
}
\label{tab1}
\[
\begin{tabular}{lcc}
\hline
\noalign{\smallskip}
 Quantity & Mode & $\sigma$ \\
\hline
\noalign{\smallskip}
$(H-K_s)_0$ & $0.14$ & $0.04$ \\
$(H-W1)_0$ & $0.18$ & $0.05$ \\
$(H-W2)_0$ & $0.18$ & $0.05$ \\
$(H-W3)_0$ & $0.17$ & $0.03$ \\
$(H-W4)_0$ & $0.24$ & $0.04$ \\
$(K_s-W1)_0$ & $0.04$ & $0.05$ \\
$(K_s-W2)_0$ & $0.04$ & $0.05$ \\
$(K_s-W3)_0$ & $0.03$ & $0.02$ \\
$(K_s-W4)_0$ & $0.10$ & $0.03$ \\
$(W1-W2)_0$ & $0.00$ & $0.06$ \\
$(W1-W3)_0$ & $-0.01$ & $0.04$ \\
$(W1-W4)_0$ & $0.06$ & $0.05$ \\
$(W2-W3)_0$ & $-0.01$ & $0.04$ \\
$(W2-W4)_0$ & $0.06$ & $0.05$ \\
$(W3-W4)_0$ & $0.07$ & $0.02$ \\
$(M_H)$ & $-1.49$ & $0.04$ \\
$(M_{K_s})$ & $-1.63$ & $0.03$ \\
$(M_{W1})$ & $-1.67$ & $0.05$ \\
$(M_{W2})$ & $-1.67$ & $0.05$ \\
$(M_{W3})$ & $-1.66$ & $0.02$ \\
$(M_{W4})$ & $-1.73$ & $0.03$ \\
\hline
\end{tabular}
\]
\end{table*}
}

\section*{ACKNOWLEDGMENTS}

This study uses the data from the Two-Micron All Sky Survey (2MASS) and the Wide-field Infrared Survey Explorer (WISE) and the resources of the 
Centre de Donn\'ees astronomiques de Strasbourg.
This work has made use of data from the European Space Agency (ESA) mission Gaia (https://www.cosmos.esa.int/gaia), 
processed by the Gaia Data Processing and Analysis Consortium (DPAC, \\
https://www.cosmos.esa.int/web/gaia/dpac/consortium). 
Funding for the DPAC has been provided by national institutions, in particular the institutions participating in the Gaia Multilateral Agreement.

\end{document}